# Negative Giant Longitudinal Magnetoresistance in NiMnSb/InSb: An interface effect

S. Gardelis,[1,*] J. Androulakis,[1,**] Z. Viskadourakis,[1,2] E.L. Papadopoulou,[1,2] and J. Giapintzakis[1, 3,+]

[1]*Institute of Electronic Structure and Laser, FORTH, P.O. Box 1527, Vassilika Vouton, 71110 Heraklion, Crete, Greece*

[2]*Department of Materials Science and Technology, University of Crete, P.O. Box 2208, 710 03 Heraklion, Crete, Greece*

[3]*Department of Mechanical & Manufacturing Engineering, University of Cyprus, P.O. Box 20537, 1678 Nicosia, Cyprus*

## ABSTRACT

We report on the electrical and magneto-transport properties of the contact formed between polycrystalline NiMnSb thin films grown using pulsed laser deposition (PLD) and *n*-type degenerate InSb (100) substrates. A negative giant magnetoresistance (GMR) effect is observed when the external magnetic field is parallel to the surface of the film and to the current direction. We attribute the observed phenomenon to magnetic precipitates formed during the magnetic film deposition and confined to a narrow layer at the interface. The effect of these precipitates on the magnetoresistance depends on the thermal processing of the system.

Recently, we reported on the successful growth of stoichiometric polycrystalline NiMnSb thin films on InSb substrates using pulsed laser deposition (PLD).[1] Our films exhibit a saturation magnetization of 4 $\mu_B$/formula unit at *5 K* and a magnetization temperature dependence that follows Bloch's $T^{3/2}$ law, consistently with the expected half-metallic behavior.

In this letter we report on the electrical and magneto-transport properties of the NiMnSb/InSb system. The contact formed between the NiMnSb film and the InSb substrate exhibits ohmic behavior. An unusual negative giant magnetoresistance (GMR) effect is found when the magnetic field is applied parallel to the in-plane current direction. A similar effect is also observed when Ni is deposited on InSb. On the other hand, no negative GMR effect is displayed when the deposited film is nonmagnetic. We argue that the GMR effect is due to magnetic precipitates formed at the interface during the growth of the magnetic films, which align their magnetic moments in the direction of the external magnetic field and thus, the spin dependent scattering of the electrons is reduced.

NiMnSb films were grown by the conventional PLD method onto heated ($T_s\sim200^oC$) *n*-type InSb (100) substrates. Further details of the growth conditions of the NiMnSb films on InSb and their structural and magnetic properties can be found elsewhere.[1] In order to investigate the electrical contact between the NiMnSb film and the InSb substrate, we performed detailed resistance and magnetoresistance (MR) measurements on both the NiMnSb/InSb system and the bulk InSb substrates. It is important to note that in the former case, the electrodes for the four-terminal measurements were placed on the top NiMnSb layer (current-in-plane, CIP, configuration). The electrical resistance was measured by the standard four-probe ac method in the temperature range *5≤T≤300 K* and in the magnetic field interval *0-7 T*.

Magnetoresistance measurements were taken for both positive and negative magnetic fields in order to eliminate effects due to probe misalignment.

As shown in Fig. 1 the I-V characteristic curve obtained at *5 K* indicates that the contact between NiMnSb and InSb is ohmic. We note that the temperature dependence of the resistance (not shown) of the *n*-type degenerate InSb substrates exhibits metallic behavior.

MR measurements of the *n*-type InSb (100) substrate revealed the presence of Shubnikov-de Haas (SdH) oscillations for magnetic fields applied both perpendicular and parallel to the crystal surface. These magneto-oscillations are due to the Landau quantization of the three-dimensional (3D) electronic density of states.[2] The MR oscillations are periodic in 1/*H* (inset (2), Fig. 1). From the period Δ*(1/H)* of the oscillations in the transverse MR we calculated the electron density, *n*, via the formula[3] $\Delta(1/H) = \frac{2e}{\hbar}(3\pi^2 n)^{-2/3}$. The period Δ*(1/H)* was found to be *0.023 T⁻¹* which corresponds to *n = 1.6 x 10¹⁸ cm⁻³* for the InSb substrate. Such a high value of *n* explains the metallic behavior of InSb and the ohmic behavior of the contact between NiMnSb and InSb. Surprisingly SdH oscillations were also observed in the MR of the NiMnSb/InSb system, i.e. following the deposition of *600 nm*-thick NiMnSb layer on top of the InSb substrate. This is a further indication of the good ohmic behavior of the NiMnSb-InSb electrical contact.

Apart from the oscillating part of the MR, the background of the MR gives some important information about the electron transport in these systems. In both InSb and NiMnSb/InSb systems the transverse MR is found to be positive and quadratic in *H*. This is expected since the applied magnetic field forces the electrons in circular orbits, which enhances the scattering of the electrons, and therefore the resistance increases with increasing magnetic field.

The longitudinal MR of the NiMnSb/InSb system, with *H* applied in-plane and parallel to the current, corresponding to zero Lorentz force on the charge carriers, exhibits the most interesting behavior. Apart from the low-temperature oscillations, there is a negative MR background which persists almost unchanged to room temperature. In Fig. 2 the ratio $\frac{\Delta R}{R}\left(=\frac{R(H)-R(H=0)}{R(H=0)}\right)$ versus *H* is plotted for *5 K, 50 K, 145 K* and *300 K*. The ratio Δ*R/R* changes from *25%* at *5 K* to *15%* at *300 K*. This is a giant magnetoresistance effect (GMR) and comparatively much larger than the negative MR which is observed in the case of the InSb substrate at low magnetic fields and low temperatures as part of an otherwise intense positive magnetoresistance background (inset, Fig. 2). In the case of the bulk InSb the small negative MR is similar to the weak disorder effect observed in *n*-type InSb by Mani *et al.*[4] However, in older literature the negative magnetoresistance observed at low magnetic fields and at low temperatures in *n*-type degenerate semiconductors is attributed to the scattering of the conduction electrons by localized spins through an *s-d* exchange interaction although magnetic impurities are not present.[5,6,7,8]

It is noteworthy that no negative GMR effect is observed both in NiMnSb bulk samples and in NiMnSb films deposited on highly resistive Si substrates[9] (in this case we measure only the magnetoresistance of the NiMnSb films) with the same *H* configuration. In addition, no negative GMR is observed for the configuration in which *H* is in-plane and perpendicular to the current.

To investigate the origin of this interesting negative longitudinal GMR effect, we performed similar MR measurements on Ni/InSb and Al/InSb systems for which Ni and Al films were deposited on heated InSb substrates at *200ºC* by PLD. Whereas we observed a negative GMR in the longitudinal MR of the magnetic Ni/InSb system, in the case of the non-magnetic Al/InSb system we observed only a positive MR without any negative MR component (Fig. 3). Moreover, we measured the longitudinal MR of a NiMnSb/InSb system, where the NiMnSb film was deposited by PLD at room temperature (Fig.4a). In this case only a small negative MR was found on an otherwise positive MR background. Interestingly, after annealing the room-temperature grown sample in a pure Ar atmosphere at *200$^o$C* for 30 minutes, the negative MR increased drastically (Fig.4b). In all of these systems we observed SdH oscillations in the MR at *5 K* and their I-V characteristics exhibited ohmic behavior.

Upon measuring the MR of the Metal/InSb systems, we observe the contribution of both the metallic film and the InSb substrate because they are both conductive. This is evident at low temperatures, where we observe the SdH oscillations originating from the substrate. In addition, the fact that the negative GMR is not observed in any of the constituents of the NiMnSb/InSb or Ni/InSb systems leads us to suggest that this effect occurs at the interface between the ferromagnetic metal and InSb. A plausible explanation for the negative GMR is that at the interface there is a thin layer of InSb containing microscopic magnetic entities (NiMnSb or Ni precipitates). Upon increasing the magnetic field, these magnetic entities gradually align their magnetic moments with the external magnetic field leading to a decrease in the spin-dependent resistance of the system. These precipitates are ablated metal particles that reach the surface of the InSb substrate with energies high enough to allow their penetration to a shallow depth. This is in agreement with our observation of a low density NiMnSb film at the interface of the NiMnSb/InSb system obtained by grazing incidence x-ray reflectometry (XRR) measurements.[10] It has been observed that either reduction of the fluence or increase of the pressure can slow down the energetic particles during PLD deposition.[11] To this end, we introduced pure Ar gas to a pressure of 0.05 mbar in order to reduce the velocities of the ablated particles. Fig. 4c indicates clearly that there is no MR in this case. Thereby this implantation-like process depends on the energy and the thermal processing of the system. Specifically, once the particles are energetic enough the temperature could promote clustering of the particles, or assist diffusion deeper into InSb, or assist in recovering the high mobility at the interface, or finally a combination of two or more of the above events could occur.

A more striking evidence of the possible existence of spin-scatterers at the interface in the NiMnSb/InSb and Ni/InSb systems is that the negative GMR effect does not occur in the Al/InSb system because Al precipitates in InSb have no magnetic moment and therefore, only the positive magnetoresistance contribution is observed.

It is worth mentioning that the observed negative GMR is similar to the one observed in Cu-Co alloys consisting of ultrafine Co precipitates in a Cu matrix when the magnetic field is applied parallel to the current.[12] It is also noteworthy that negative magnetoresistance has been observed in erbium-, yttrium- and europium-doped InSb films and has been attributed to scattering of conduction electrons by the magnetic spins of the afore mentioned rare earth atoms.[13,14,15]

Furthermore, the quadratic dependence of the negative GMR on the magnetic field observed in our samples for magnetic fields up to 2 $T$ is in agreement with theoretical calculations of negative magnetoresistance due to localized spins.[16]

The magnetic precipitates at the interface can act as spin scatterers and as a result may obstruct spin injection from the magnetic metal into InSb. Therefore, great care must be taken during the formation of the magnetic film in order to avoid the presence of such spin scatterers at the interface.

In conclusion an unusual negative GMR effect persisting at room temperature is observed in ferromagnetic metal/InSb systems when the magnetic field is applied in-plane and parallel to the current. This effect is attributed to magnetic precipitates formed at the interface between the ferromagnetic metal film and the InSb substrate.

This work was supported by the EU project FENIKS (G5RD-CT-2001-00535). The authors gratefully acknowledge Dr. P.D. Buckle for supplying the InSb substrates. We would also like to thank Dr. G. Kastrinakis for useful discussions.

# REFERENCES

[+]Electronic address: giapintz@ucy.ac.cy

[*]Present address: Institute of Microelectronics, NCSR "Demokritos", 153 10 Aghia Paraskevi, Attiki, Greece

[**]Present Address: Department of Chemistry, Michigan State University, East Lansing, MI 48824, USA

[1] S. Gardelis, J. Androulakis, J. Giapintzakis, O. Monnereau and P.D. Buckle, Appl. Phys. Lett. **85**, 3178 (2004)

[2] D.G. Seiler, Phys. Letters **31A**, 309 (1970); G. Bauer and H. Kahlert, J. Phys. C: Solid State Phys. **6**, 1253 (1973)

[3] D. Jena, S. Heikman, J.S. Speck, A. Gossard, U.K. Mishra, A. Link and O. Ambacher, Phys. Rev. B **67**, 153306 (2003)

[4] R.G. Mani, L. Ghenim and J.B. Choi, Phys. Rev. B **43**, 12630 (1991)

[5] Y. Katayama and S. Tanaka, Phys. Rev. **153,** 873 (1967)

[6] R.P. Khosla and J.R. Fischer, Phys. Rev. B **2**, 4084 (1970)

[7] R.P. Khosla and J.R. Fischer, Phys. Rev. B **6**, 4073 (1972)

[8] G. Bauer and H. Kahlert, Phys. Rev. B **5**, 566 (1972)

[9] W.R. Branford, S.K. Clowes, M.H. Syed, Y.V. Bugoslavsky, S. Gardelis, J. Androulakis, J. Giapintzakis, C.E.A. Grigorescu, A.V. Berenov, S.B. Roy and L.F. Cohen, Appl. Phys. Lett. **84**, 2358 (2004)

[10] S. Rai, M.K. Tiwari, M.H. Modi, G.S. Lodha, R.V. Nandedkar, S. Gardelis, Z. Viskadourakis, J. Giapintzakis, S.B. Roy and P. Chaddah, unpublished

[11] K. Sturm and Hans-Ulrich Krebs, J. Appl. Phys. **90**, 1061 (2001)

[12] A.E. Berkowitz, J.R. Mitchell, M.J. Carey, A.P. Young, S. Zhang, F.E. Spada, F.T. Parker, A. Hutten and G. Thomas, Phys. Rev. Lett. **68**, 3745 (1992); J.Q. Xiao, J.S. Jiang and C.L. Chien, Phys. Rev. Lett. **68**, 3749 (1992)

[13] J. Heremans, D.L. Partin, D.T. Morelli and C.M. Thrush, J. Vac. Sci. Technol. B **10**, 659 (1992)

[14] J. Yang, J. Heremans, D.L. Partin, C.M. Thrush and R. Naik, J. Appl. Phys. **83**, 2041 (1997)

[15] D.T. Morelli, D.L. Partin, J. Heremans, C.M. Thrush, J. Vac. Sci. Technol. B **10**, 110 (1992)

[16] M.R. Boon, Phys. Rev. B **7**, 761 (1973); Y. Toyozawa, J. Phys. Soc. Jap. **17**, 986 (1962)

# FIGURE CAPTIONS

**FIG. 1.** I-V characteristic for the NiMnSb/InSb system at *5 K*. Inset 1: Schematic of the set up used for the I-V measurements. Inset 2: The oscillating part of the transverse MR as a function of 1/H for the InSb substrate at *5 K* (after removing the background by fitting it with a polynomial).

**FIG. 2.** Plots of the ratio Δ*R/R* as a function of the magnetic field *(H)* at *5 K, 50 K, 145 K* and *300 K*. H is applied parallel to the current direction. The curves are shifted along the y-axis for clarity. Inset: Plot of the ratio Δ*R/R* versus *H* at *5 K* for the InSb substrate with the same *H* configuration.

**FIG. 3.** Plots of the ratio Δ*R/R* as a function of the magnetic field *(H)* for (a) NiMnSb/InSb, (b) Ni/InSb, and (c) Al/InSb at *5 K*. *H* is applied parallel to the current direction. The curves are shifted along the y-axis for clarity.

**FIG. 4.** Plots of the ratio Δ*R/R* versus *H* at *300 K* for the NiMnSb/InSb system: (a) the NiMnSb film is deposited by PLD at room temperature, (b) following post-deposition annealing at 200°C in Ar and (c) the NiMnSb film is deposited at 200°C at a pressure of 0.05 mbar of Ar. *H* is applied parallel to the current direction. The curves are shifted along the y-axis for clarity.

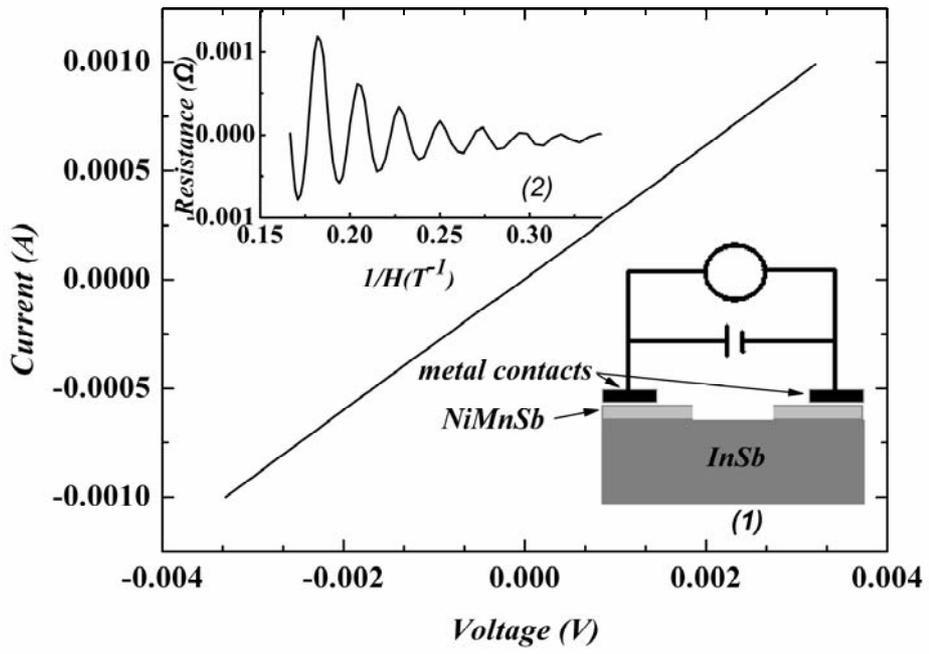

**FIG. 1.** S. Gardelis et al., APL

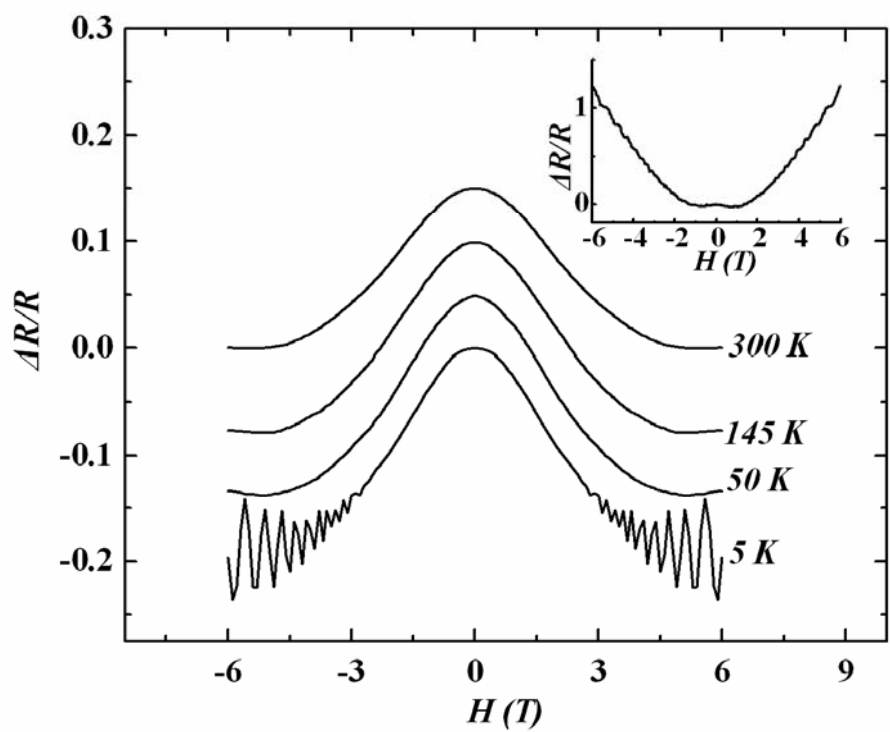

**FIG. 2.** S. Gardelis et al., APL

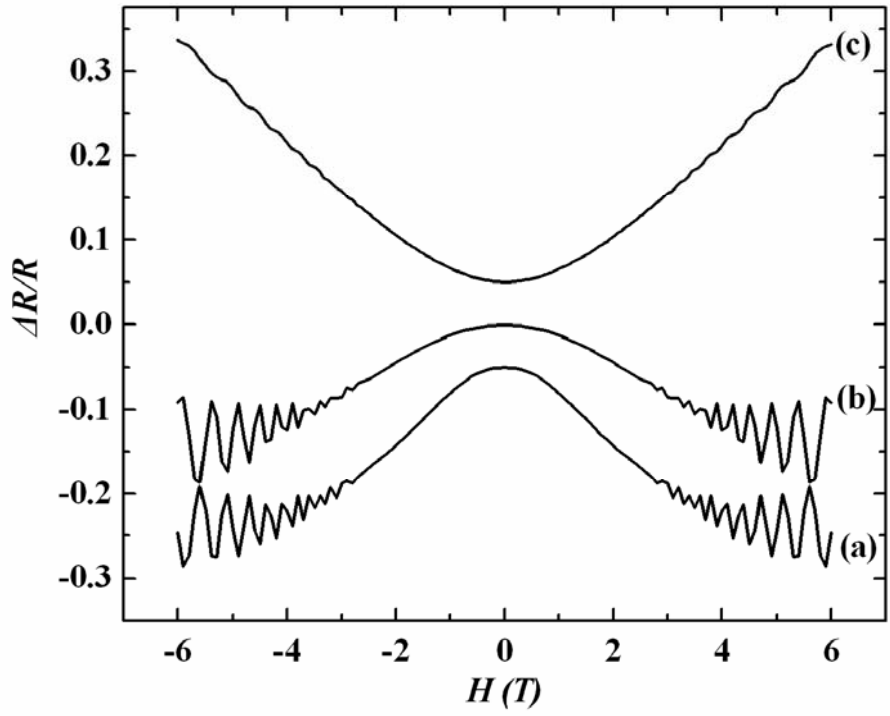

**FIG. 3.** S. Gardelis et al., APL

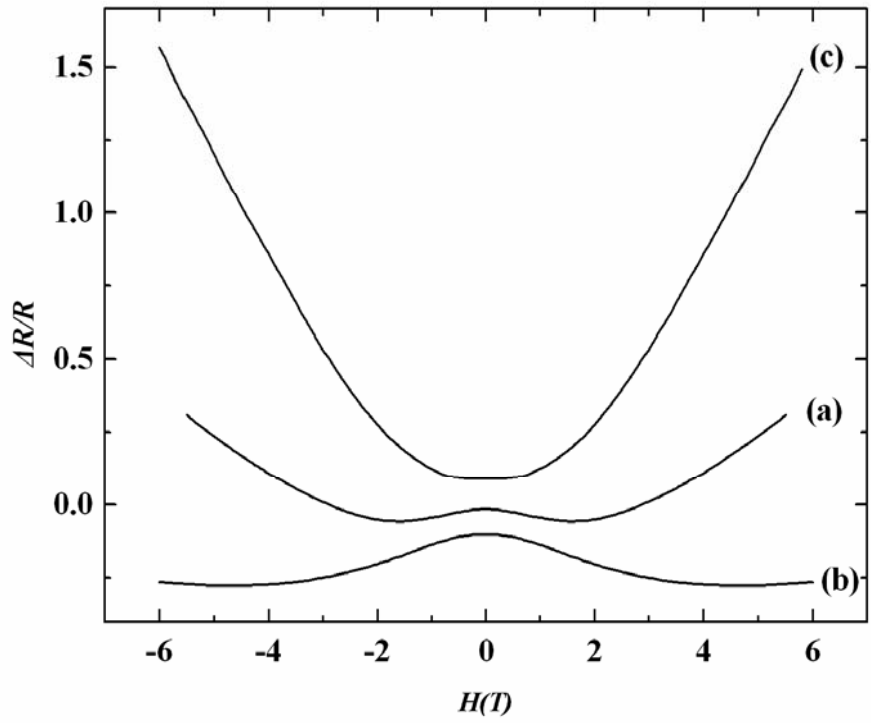

**FIG. 4.** S. Gardelis et al., APL